\documentclass[nofootinbib,aps,reprint,amsmath,amssymb]{revtex4-1} 
\usepackage{hyperref}
\usepackage{changes}
\usepackage[section]{placeins}
\usepackage[titletoc]{appendix}
\usepackage{xcolor}
\usepackage{dsfont}
\usepackage{bm}
\usepackage{mathtools}
\usepackage{enumerate}
\usepackage{braket}
\usepackage{scalerel}
\DeclareMathAlphabet{\mathpzc}{OT1}{pzc}{m}{it}

\newcommand{\sayy}[1]{`#1'}
\DeclarePairedDelimiter\abs{\lvert}{\rvert}%
\providecommand{\href}[2]{#2} 

\def\be{\begin{equation}}
\def\ee{\end{equation}}
\def\bea{\begin{eqnarray}}
\def\eea{\end{eqnarray}}

\def\sig{\sigma}

\def\la{\langle}
\def\ra{\rangle}
\def\Eu{ \mathfrak{H} }

\def\obs{\mathcal{O}}
\def\emi{\mathcal{E}}

\providecommand{\href}[2]{#2}
\def\MNRAS#1{{\em Mon.\ Not.\ R.\ Astr.\ Soc.}\ {\bf#1}}

\def\apj{{Astrophys. J.} }

\def\jcap{{J. Cosmol. Astropart. Phys.} }

\definecolor{MyB}{rgb}{0.1,0.1,1.0}

\begin{document}
\title{Differential age observations and their constraining power in cosmology} 
\author{Asta~Heinesen}
\email{asta.heinesen@nbi.ku.dk}
\affiliation{Niels Bohr Institute, Blegdamsvej 17, DK-2100 Copenhagen, Denmark}

\begin{abstract} 
We derive the differential age signal valid for cosmic chronometers (passively evolving galaxies) in any space-time that satisfies the following assumptions: (i) The space-time has a metric with Lorentzian signature and the connection is the Levi-Civita connection;  
 (ii) the cosmic chronometers are collectively well approximated as a geodesic and irrotational congruence of time-like worldlines in the space-time; (iii) light travels on null geodesics and caustics on the observer's past light cone can be ignored; (iv) the space-time is cosmological, meaning that isotropic and positive expansion degrees-of-freedom dominate over anisotropic and negative expansion degrees-of-freedom when viewed on sufficiently large scales in the frame of the cosmic chronometers.  

The main result of the paper is an expression for the differential age signal that is written in terms of line-of-sight averages of the expansion rate along individual null lines, thus providing a kinematic interpretation of the differential age signal applicable to cosmological space-times satisfying (i)--(iv). 
We explain how this result indicates that the differential age signal is a robust probe of the volume-average expansion rate in very general statistically homogeneous and isotropic space-time scenarios where other probes of the volume-average expansion rate tend to yield biased results.  
We argue that this unique property of the differential age signal makes it an ideal measurement for constraining the expansion history model-independently.

\end{abstract}

\maketitle

\section{Introduction}
%%%%%%%%%%%%%%%%%%%%%%%%%%%%%%%%%%%%%%%%%%%%%%%%%%%%%%%%%%%%%%%%%%% 
Since the first observational indications of cosmic expansion of space \cite{Lemaitre:1927,Hubble:1929}, our determination of the expansion history of the Universe has evolved from being an extrapolative mapping of a sparse number of astrophysical sources to being a precise measurement involving large cosmological datasets.  
Conventional late-Universe probes of the expansion history include the baryon acoustic oscillation scale in the galaxy distribution \cite{Bassett:2009mm} 
and the calibrated distances to supernovae \cite{Riess:2022oxy}.  
These probes constrain cosmological distances to the (mean) redshifts of the samples from which the local expansion rate can be reconstructed 
within a parametrisation of the cosmology, which is in practice most often chosen to be the $\Lambda$ cold dark matter ($\Lambda$CDM) model or a Friedmann-Lema\^{\i}tre-Robertson-Walker (FLRW) cosmography. 

Given the seemingly persistent tensions in model parameters that are present when fitting the $\Lambda$CDM model to different cosmological datasets \cite{Freedman:2017yms,Bullock:2017xww,Riess:2019cxk,Perivolaropoulos:2021jda,Peebles:2022akh}, it is worthwhile to consider complementary probes of the expansion rate of the Universe. 
The measurement of differential ages \cite{Jimenez:2001gg} is one such a probe. 
Differential age measurements rely on the identification of cosmic chronometers\footnote{Passively evolving galaxies, which are galaxies that have finished their star forming activities at early cosmic times are in practice used as chronometers \cite{Moresco:2018xdr}.}, which are astrophysical objects with standardisable ages, and their redshifts. 
Cosmic chronometers allow to identify the relative increments in ages, $\delta \tau$, and redshifts, $\delta z$, across chronometers, which in turn allow to compute the differential age signal: $\delta z /\delta\tau$. 
Measurements of the differential age signal have so far been carried out in the interval \cite{Moresco:2015cya} $0.1 \lesssim z \lesssim 2$. 
Due to the systematic uncertainties in the calculation of the relative ages as well as statistical errors, typical uncertainties for current estimates of the differential age signal with, e.g., the Sloan Digital Sky Survey are of the order $\sim 10$ percent. 
These uncertainties will be reduced with data observed by Euclid, which is expected to yield an independent measurement of $4$ percent uncertainty \cite{Moresco:2022phi}, and total errors may be brought down to below 1 percent in the future if models of the stellar population synthesis are improved sufficiently \cite{Moresco:2020fbm}.

In FLRW universe models (and for infinitesimal separations of the chronometers in age relative to the Hubble length scale) we have that the differential age signal reduces to $\delta z /\delta\tau = - (1+z) H$, where $H$ is the Hubble parameter as evaluated at the mean redshift of the sample.  
The majority of investigations of differential ages in the literature have been aimed at constraining cosmological parameters within the spatially-flat $\Lambda$CDM model and its extensions with alternative dark energy equations of state, e.g., \cite{Jimenez:2003iv,Simon:2004tf,2010JCAP...02..008S,2012ApJ...758..107L}. 
Cosmic chronometer data have also been used for consistency testing the FLRW metric as a global cosmological model \cite{Montanari:2017yma}, as a means to address tensions relating to the Hubble expansion rate \cite{Busti:2014dua,Gomez-Valent:2018hwc} and spatial curvature \cite{Vagnozzi:2020dfn,Favale:2023lnp}, for constraining a class of  Lema\^{\i}tre-Tolman-Bondi models \cite{2012ApJ...748..111W}, and as a test of cosmological backreaction effects \cite{Koksbang:2021qqc}.  

The differential nature of the measurement makes it particularly interesting as a model-independent measurement of the expansion rate of the Universe, and a 
few of the above-mentioned analyses \cite{Montanari:2017yma,Favale:2023lnp} have used model-independent re-construction techniques for the \sayy{Hubble parameter} beyond FLRW space-times, defined from the differential age signal as \sayy{$H$} $:= - \delta z /\delta\tau  / (1+z)$. 
It is of course not clear what a Hubble parameter means once going beyond the FLRW universe models, and the above definition may be thought of as one possible empirical generalisation of the Hubble parameter. 
While measurements of the differential $\delta z /\delta\tau$ can in principle be carried out in any imagined universe-model that allows the existence of cosmic chronometers, the geometrical interpretation of such measurements may be non-trivial, and it is therefore not \emph{a priori} given that $- \delta z /\delta\tau  / (1+z)$ is directly interpretable as measuring the (average) expansion rate of space in non-trivial space-time scenarios. 

It is the purpose of this paper to make a geometrical interpretation of the differential age signal possible in very general space-time scenarios. 
This is done by formulating the differential age signal in terms of kinematic variables associated with the congruence of cosmic chronometers that is assumed to exist in the space-time. 
We first consider a simplistic measurement of the differential age signal based on two cosmic chronometers, and then generalise this to a realistic statistical inference of the signal based on multiple chronometers.  
Apart from it being of theoretical interest to examine the nature of the differential age signal in so-far unexplored space-time solutions, the derived expressions 
provide a framework to incorporate the impacts of cosmic structures on the differential age signal in exact, perturbative, or numerical models of cosmic structures.

\vspace{5pt} 
\noindent
\underbar{Notation and conventions:}
Units are used in which $c=1$. Greek letters $\mu, \nu, \ldots$ label spacetime
indices in a general basis, and repeated indices are summed over. 
The signature of the space-time metric $g_{\mu \nu}$ is $(- + + +)$ and $\nabla_\mu$ is the covariant derivative defined by the Levi-Civita connection. 
Round brackets $(\, )$ around indices denote symmetrisation in the indices, and square brackets $[\, ]$ denote anti-symmetrisation. 
Bold notation $\bm V$ for the basis-free representation of vectors $V^\mu$ is used occasionally. 
Latin letters $i, j, \ldots$ are used to label objects (cosmic chronometers) in the analysis. 
%%%%%%%%%%%%%%%%%%%%%%%%%%%%%%%%%%%%%%%%%%%%%%%%%%%%%%%%%%%%%%%%%%%%
%%%%%%%%%%%%%%%%%%%%%%%%%%%%%%%%%%%%%%%%%%%%%%%%%%%%%%%%%%%%%%%%%%%  

\section{Assumptions and setup}
\label{sec:setup}
%%%%%%%%%%%%%%%%%%%%%%%%%%%%%%%%%%%%%%%%%%%%%%%%%%%%%%%%%%%%%%%%%%%  
We assume a space-time with a metric of Lorentzian signature 
that admits the presence of an irrotational congruence of worldlines generated by a geodesic 4--velocity field $\bm u$. The geodesic and irrotational requirements are equivalent to setting the 4-acceleration, $\bm a$, and the vorticity tensor, $ \omega_{\mu \nu}$, to zero: 
\bea
\label{def:geodesicirr} 
a^\mu \equiv u^\nu \nabla_\nu u^\mu = 0 \, \qquad  \omega_{\mu \nu} \equiv h_{  \nu  }^{\, \beta}  h_{  \mu }^{\, \alpha }\nabla_{  [ \beta}u_{\alpha ] } = 0 \, , 
\eea 
where $h_{ \mu }^{\; \nu } \equiv u_{ \mu } u^{\nu } + g_{ \mu }^{\; \nu } $ is the spatial projection tensor in the rest frame of the 4--velocity field $\bm u$. 
We assume that $\bm u$ defines the frame of the cosmic chronometers present in the space-time.   
The geodesic and irrotational properties (\ref{def:geodesicirr}) allow us to write $\bm u$ in terms of a gradient of a scalar function 
\bea
\label{def:ugradient} 
u_\mu = - \nabla_\mu \tau    \, , 
\eea 
where $\tau$ is a proper time scalar of the congruence of worldlines, in that it satisfies $u^\mu \nabla_\mu \tau =1$. 
The space-time function $\tau$ constitutes a preferred choice of proper time scalar\footnote{The proper time condition $u^\mu \nabla_\mu T =1$ is invariant under transformations $T \mapsto T + X$ of the proper time function, where the translation $X$ satisfies $u^\mu \nabla_\mu X = 0$. This freedom of translation amounts to the freedom of choosing a 3-dimensional surface of synchronisation of the proper time parameters of the individual worldlines. The condition  $T=\tau$, $u_\mu = - \nabla_\mu \tau$, which is viable only in the case where $\bm u$ is irrotational, can thus be thought of as a gauge fixing of the synchronisation.}, since constant level hyper-surfaces of $\tau$ are simultaneously hyper-surfaces orthogonal to $\bm u$. 
The proper time scalar function $\tau$ measures the age of the cosmic chronometers, and we shall thus refer to $\tau$ as the age function. 
We can write the expansion variables in the frame of the congruence of cosmic chronometers
\bea
\label{def:expu}
&& \theta \equiv \nabla_{\mu}u^{\mu} \, ,  \quad \sig_{\mu \nu} \equiv h_{ \la \nu  }^{\, \beta}  h_{  \mu \ra }^{\, \alpha } \nabla_{ \beta }u_{\alpha  }  \, 
\eea  
where $\theta$ is the isotropised expansion rate and $\sig_{\mu \nu}$ is the shear rate. 
We do not assume any specific parametrisation for these variables, but we do assume that $\theta$ is positive in most regions and that $\sig_{\mu \nu}$ is subdominant to $\theta$ in the majority of the space-time volume probed, so that the model universe can be said to be overall expanding in the frame of the chronometers. 
This requirement in practice implies a cosmological space-time where the regions with bound structures and collapse of matter are subdominant in volume to the regions that are expanding. 

Let us consider an observer who belongs to the congruence of cosmic chronometers\footnote{If the observer does not belong to the congruence, but has a relative velocity to it, the redshift measurements are subject to a boost correction.}, making observations from the space-time point $\obs$. 
We now consider a set of $N$ astrophysical sources comoving with $\bm u$ with worldlines crossing the past null cone of the event of observation $\obs$. 
For the $i$'th astrophysical source, $1 \! \leq \! i \! \leq \! N$, we denote the event of emission $\emi_i$. 
We let the 4--momentum $\bm k_i$ be generator of the null geodesic path $\gamma_i$ connecting the point of emission $\emi_i$ with the point of observation $\obs$, and we assume that there is no caustics involved so that this path is unique. 
We define the associated affine parameter $\lambda_i$ of the null ray through the transport rule $k_i^\mu \nabla_\mu \lambda_i = 1$. 
It will furthermore be convenient to decompose the 4-momentum of the null ray in terms of $\bm u$ and a spatial unit vector $\bm e_i$ in the following way 
\bea
\label{kdecomp}
k_i^\mu = E_i(u^\mu - e_i^\mu) \,  ,  \qquad e_i^\mu u_\mu \equiv 0 \, , \quad  E_i \equiv - k_i^\mu u_\mu   \, ,  
\eea  
where $E_i$ is the energy of the photon as measured by an observer comoving with $\bm u$, with evolution along the null ray 
\bea
\label{def:Eevi}
 \Eu_i \equiv - \frac{ k_i^{\mu}\nabla_{\mu} E_i }{E_i^2} =  \frac{1}{3}\theta  + e_i^\mu e_i^\nu \sigma_{\mu \nu}   \, . 
\eea  
The spatial unit vector $\bm e_i$ represents the direction of incoming light as seen by an observer comoving with the frame of the cosmic chronometers. The function $\Eu_i$ has interpretation as measuring the expansion of space in the frame of the cosmic chronometers along the direction $\bm e_i$. 

\section{Differential age signal}
\label{sec:diffage}
The redshift function evaluated at a given event\footnote{In practice, the observer measures the redshift to the event $\emi_i$. However, it is useful as an intermediate step in the calculation to define the redshift along the null geodesic line segment.} along the null geodesic path $\gamma_i$ is defined in the following way
\bea
\label{def:z}
z_i \equiv \frac{E_i }{E_i \rvert_{\obs}} - 1    \, , 
\eea  
where $\rvert_{\obs}$ denotes evaluation at the point of observation.      
We can write the evolution of $z_i$ along the null ray in the following way 
\bea
\label{def:zev}
 k_i^{\mu}\nabla_{\mu} z_i = - E_i (1+z_i)  \Eu_i \, , 
\eea  
which can be obtained directly from the energy propagation equation (\ref{def:Eevi}). 
Using (\ref{def:zev}), we may rewrite the redshift function (\ref{def:z}) as 
\bea
\label{def:zint}
z_i  =  e^{\int^{\lambda_i \rvert_{\obs}}_{\lambda_i}  d\lambda_i' \, E_i  \Eu_i (\lambda_i') }  - 1 =   e^{\int^{\tau \rvert_{\obs} }_{\tau(\lambda_i)}  d\tau \,  \Eu_i (\tau) }  - 1 \, ,
\eea  
where, in the last equality, we have used that the age function $\tau$ can be used as a parameter on any individual null ray 
with Jacobian $$d\tau / d\lambda_i \equiv k_i^\mu\nabla_\mu \tau = E_i \, ,$$
which follows from using (\ref{def:ugradient}) in the definition of $E_i$ in (\ref{kdecomp}).  
For integrals of the form $\int^{\tau_B}_{\tau_A}  d\tau \,  \Eu_i (\tau)$ it is implicit that the domain of integration is the section of $\gamma_i$ defined by the age function $[\lambda_i(\tau_A) ,  \lambda_i(\tau_B)]$.

Let us consider two astrophysical sources with points of emission $\emi_i$ and $\emi_j$ and associated null rays generated by $\bm k_i$ and $\bm k_j$. 
We define the mean redshift and mean age of the sources as $z_{ij} \equiv   ( z_{\scaleto{\emi_i \mathstrut}{5.5pt}}  + z_{\scaleto{\emi_j \mathstrut}{5.5pt}} )/2$ and  $\tau_{ij} \equiv   ( \tau_{\scaleto{\emi_i \mathstrut}{5.5pt}}  + \tau_{\scaleto{\emi_j \mathstrut}{5.5pt}} )/2$, where we have used the short hand notations $z_{\scaleto{\emi_i \mathstrut}{5.5pt} } \equiv z_i \rvert_{\emi_i}$ and $\tau_{\scaleto{\emi_i \mathstrut}{5.5pt}} \equiv \tau \rvert_{\emi_i}$. 
We define the distance in redshift of the sources away from the mean $\delta z_{i} \equiv z_{\scaleto{\emi_i \mathstrut}{5.5pt} } - z_{ij}$ and $\delta z_{j} \equiv z_{\scaleto{\emi_j \mathstrut}{5.5pt} } - z_{ij}$, and similarly we define the distance in age function away from the mean $\delta \tau_{i} \equiv \tau_{\scaleto{\emi_i \mathstrut}{5.5pt}}  - \tau_{ij}$ and $\delta \tau_{j} \equiv \tau_{\scaleto{\emi_j \mathstrut}{5.5pt}}  - \tau_{ij}$. 
We assume that the mean redshift, $z_{ij}$, is realised at least once for the redshift function along each of the null rays\footnote{This is in practice satisfied for the assumed cosmological space-times outlined in the assumption section \ref{sec:setup}, where space is expanding on large scales in the frame of the cosmic chronometers.}, i.e., there are values of the affine parameters  ${\lambda}_i = \bar{\lambda}_i$ and ${\lambda}_j = \bar{\lambda}_j$ along $\gamma_i$ and $\gamma_j$ respectively, such that $z_i(\bar{\lambda}_i) = z_{ij}$ and $z_j(\bar{\lambda}_j) = z_{ij}$.

With the above definitions and the identity \ref{def:zint}, we may write the redshift $z_{\scaleto{\emi_i \mathstrut}{5.5pt} }  \equiv  E_i \rvert_{\scaleto{\emi_i \mathstrut}{5.5pt} } / E_i \rvert_{\obs} - 1$ as 
\bea
\label{def:zint1} 
\hspace{-0.5cm} z_{\scaleto{\emi_i \mathstrut}{5.5pt} }  +  1  &&=  (z_{ ij } +1 ) \,  \exp \left( \int^{ \tau( \bar{\lambda}_i )  }_{\tau_{\scaleto{\emi_i \mathstrut}{5.5pt}}  } \!\!\! \! d\tau \,  \Eu_i (\tau)  \right)  \\ 
  &&  \approx   (z_{ ij } +1 ) \,  \left( 1 + \int^{ \tau( \bar{\lambda}_i )  }_{\tau_{\scaleto{\emi_i \mathstrut}{5.5pt}}  } \!\!\! \! d\tau \,  \Eu_i (\tau)  \right)  \, ,   
\eea  
where the approximation in the second line comes from linearising the expression in $\int^{ \tau( \bar{\lambda}_i )  }_{\tau_{\scaleto{\emi_i \mathstrut}{5.5pt}}  } \!\!  d\tau \,  \Eu_i (\tau)$, assuming that this is small. 
This is a good approximation when $\delta z_{i} \ll 1$. 
Similarly we may write $z_{\scaleto{\emi_j \mathstrut}{5.5pt} }  \equiv  E_j \rvert_{\scaleto{\emi_j \mathstrut}{5.5pt} } / E_j \rvert_{\obs} - 1$ as linearised in $\int^{ \tau( \bar{\lambda}_j )  }_{\tau_{\scaleto{\emi_i \mathstrut}{5.5pt}}  } \!\!  d\tau \,  \Eu_j (\tau)$  as 
\bea
\label{def:zint2}
 z_{\scaleto{\emi_j \mathstrut}{5.5pt} }  +  1   \approx    (z_{ ij } +1 )  \left( 1 + \int^{ \tau( \bar{\lambda}_j )  }_{\tau_{\scaleto{\emi_j \mathstrut}{5.5pt}}  } \!\!\! \! d\tau \,  \Eu_j (\tau)  \right)   \,  . 
\eea  
We can now subtract $z_{ ij } +1$ from \eqref{def:zint1} to arrive at the following equation 
\bea
\label{def:zint1dif} 
\hspace{-0.5cm} - \frac{ \frac{ \delta z_{i} }{ \delta \tau_{i}  } }{  z_{ ij } +1  } &&  \approx  
  \braket{\Eu_i}   \, ,   
\eea  
where we have defined the average expansion variable 
\bea
\label{def:havdef} 
\hspace{-0.5cm} \braket{\Eu_i} \equiv  \frac{ \int_{ \tau( \bar{\lambda}_i )  }^{\tau_{\scaleto{\emi_i \mathstrut}{5.5pt}}  } \!\!  d\tau \,  \Eu_i (\tau) }{ \delta \tau_{i}  }  \, .   
\eea   
The expression \eqref{def:zint1dif} gives the differential age signal estimate for the pair of chronometers in terms of the projected expansion rate of the congruence of cosmic chronometers along $\gamma_i$. 
A similar expression can be derived for $\frac{ \delta z_{j} }{ \delta \tau_{j}  }$ by an exchange of indices $(i \leftrightarrow j)$ in \eqref{def:zint1dif} and \eqref{def:havdef}. 
The final expression \eqref{def:zint1dif} for the differential age signal have close resemblance to the FLRW expression \cite{Moresco:2015cya} $-\delta z /\delta\tau / (1+z) =  H$, when $\braket{\Eu_i}$ is interpreted as an effective Hubble parameter. 
Indeed, for long distances of light propagation relative to a suitable homogeneity scale, the light ray adapted expansion variable $\Eu$ can with a high level of accuracy be replaced by (one third of) the volume-average expansion rate in the matter frame for a large class of non-trivial cosmological solutions \cite{Lavinto:2013exa,Koksbang:2017arw,Koksbang:2020qry}. 

A suitable weighting of \eqref{def:zint1dif} and the analogous expression for the j'th source may thus probe the average expansion rate of the Universe if the length intervals $c \abs{ \delta \tau_{i} }$ and $c \abs{ \delta \tau_{j}}$ are comparable or larger than an approximate homogeneity scale, but still small enough for the above linearisations in $\int^{ \tau( \bar{\lambda}_i )  }_{\tau_{\scaleto{\emi_i \mathstrut}{5.5pt}}  } \!\!  d\tau \,  \Eu_i (\tau)$ and $\int^{ \tau( \bar{\lambda}_j )  }_{\tau_{\scaleto{\emi_j \mathstrut}{5.5pt}}  } \!\!  d\tau \,  \Eu_j (\tau)$ to be accurate. 
In practice, estimates of the differential age signal are made based on statistical analyses of large catalogues of galaxies. 
In the below we shall analyse a realistic estimator for the differential age signal.

\section{Averaging over sources}
\label{sec:obs} 
In practice, the differential age signal is inferred from many cosmic chronometers in the form of passively evolving galaxies in the upper envelope of ages as a function of redshift \cite{2010JCAP...02..008S}. 
Consider a setup where we have a large number, $N$, of such cosmic chronometers labelled by $i = 1, ... , N$. 
We let $\bar{z}$ be a suitably defined mean redshift of the sample of chronometers and $\bar{\tau}$ be a corresponding mean age of the sample\footnote{The parameters $\bar{z}$ and $\bar{\tau}$ may be left free in the likelihood function/ chi-square statistic and determined simultaneously with the differential age signal in the statistical analysis.}.  
We let $\delta z_i \equiv z_i - \bar{z}$ and $\delta \tau_i \equiv \tau_i - \bar{\tau}$ be the redshift and age increments of each source relative to the mean. 

We want to consider the likelihood of these measured redshift and age increments of the cosmic chronometers given a model.  
Suppose that the cosmic chronometers cover a cosmological volume of space over which space is expanding, but which are still closely enough placed in terms of their redshift and proper ages, such that the increments in $\delta z_i$ and $\delta \tau_i$ can indeed be considered infinitesimal in a cosmological context, such that a linear relation can be assumed to hold between them\footnote{This approach was for instance taken in \cite{Jimenez:2023flo} in order to estimate the differential age signal in a model-independent way.}. 
We assume that the residuals $( \delta \tau_i  -  \frac{\delta T}{ \delta Z }   \, \delta z_i )$ are distributed according to a multivariate Gaussian distribution with covariance matrix $\text{COV}$, such that 
\bea
\label{def:chi}
\hspace{-0.5cm}  \chi^2 = \sum^N_{i=1} \! \sum^N_{j=1} (\text{COV}^{-1})_{ij} \!  \left( \! \delta \tau_i  - \frac{\delta T}{ \delta Z } \delta z_i  \! \right) \! \left( \! \delta \tau_j  - \frac{\delta T}{ \delta Z } \delta z_j  \! \right) \,  
\eea  
is chi-squared distributed. 
In \eqref{def:chi}, we view the redshift as the input parameter and the age as the response variable in the parametrisation of the age--redshift curve that we are constraining with the chi-square statistic. 
This is because it is the age measurement that comes with the largest uncertainty in practice, whereas the redshift the redshift is usually very precisely determined. 
In a frequentist approach we may minimise $\chi^2$ to obtain the best fit differential age signal 
\bea
\label{def:chidif}
\hspace{-0.2cm}  && \frac{\partial \chi^2}{\partial  ( \frac{\delta T}{ \delta Z } )  } \Big\rvert_{\frac{\delta T}{ \delta Z } = \hat{\frac{\delta T}{ \delta Z } }} =  0  \nonumber \\ 
&&  \Leftrightarrow \hat{\frac{\delta T}{ \delta Z } } =  \frac{ \sum^N_{i=1}  \sum^N_{j=1} (\text{COV}^{-1})_{ij}  \delta \tau_i \delta z_j }{  \sum^N_{i=1}  \sum^N_{j=1} (\text{COV}^{-1})_{ij}    \frac{ \delta z_i }{\delta \tau_i}  \delta \tau_i  \delta z_j   } \, , 
\eea 
i.e., we have that the result for the best fit slope $\hat{\frac{\delta T}{ \delta Z } }$ is given by weighted sums of the individual $\delta \tau_i$ and $\delta z_i$ measurements. 

We may now analyse the meaning of the estimate in \eqref{def:chidif} in terms of the geometry of the space-time. 
Going through the same calculation as in section \ref{sec:diffage}, we may derive to leading order in $\int_{ \tau( \bar{\lambda}_i )  }^{\tau_{\scaleto{\emi_i \mathstrut}{5.5pt}}  } \!\!  d\tau \,  \Eu_i (\tau)$ that 
\bea
\label{def:zint1difagain}
\hspace{-0.5cm} \frac{ \delta z_{i} }{ \delta \tau_{i}  }&&  \approx   
-  (\bar{z} +1 )   \braket{\Eu_i}   \, ,   
\eea  
with 
\bea 
\label{def:havdefagain} 
\hspace{-0.5cm} \braket{\Eu_i} \equiv  \frac{ \int_{ \tau( \bar{\lambda}_i )  }^{\tau_{\scaleto{\emi_i \mathstrut}{5.5pt}}  } \!\!  d\tau \,  \Eu_i (\tau) }{ \delta \tau_{i}  }  \, , 
\eea  
where $\bar{\lambda}_i$ is defined as a value of the affine parameter $\lambda_i$ along $\gamma_i$ satisfying $z_i(\bar{\lambda}_i) = \bar{z}$. 
Substituting \eqref{def:zint1difagain} in \eqref{def:chidif}, we get that the best fit value \eqref{def:chidif} corresponds to the quantity 
\bea
\label{def:chidiftheory}
\hspace{-0.4cm}  \hat{\frac{\delta T}{ \delta Z }} \approx   -  \frac{1}{\bar{z} +1}  \frac{ \sum^N_{i=1}  \sum^N_{j=1} (\text{COV}^{-1})_{ij}     \delta \tau_i \delta z_j }{  \sum^N_{i=1}  \sum^N_{j=1} (\text{COV}^{-1})_{ij}    \braket{\Eu_i} \delta \tau_i \delta z_j   } \, , 
\eea 
where again, the approximation holds to leading order in $\int_{ \tau( \bar{\lambda}_i )  }^{\tau_{\scaleto{\emi_i \mathstrut}{5.5pt}}  } \!\!  d\tau \,  \Eu_i (\tau)$. 
Inverting \eqref{def:chidiftheory}, we have 
\bea
\label{def:chidiftheory_inv}
\hspace{-0.3cm} -  \frac{ \, \left( \hat{\frac{\delta T}{ \delta Z }} \right)^{ \! \! -1} }{\bar{z} +1}   \approx  \frac{  \sum^N_{i=1}  \sum^N_{j=1} (\text{COV}^{-1})_{ij}    \braket{\Eu_i} \delta \tau_i \delta z_j    }{   \sum^N_{i=1}  \sum^N_{j=1} (\text{COV}^{-1})_{ij}     \delta \tau_i \delta z_j    } \, , 
\eea 
to the same accuracy of approximation. 
The right hand side of \eqref{def:chidiftheory_inv} consists of a double average of the expansion variable $\Eu_i$ along the null rays. The inner averages, $\braket{\Eu_i}$, are taken over the individual null rays, $\gamma_i$, whereas the outer average is taken over the cosmic chronometers of the sample. 
The latter average has the usual weighting by the inverse covariance and an additional weighting by $\delta \tau_i \delta z_j$, which may be interpreted as a squared path length of integration along the ray $\gamma_i$, thus upweighting the cosmic chronometers that are the furthest away from from the center of the survey. 

What happens when we are probing a large volume with the survey such that second order correction in $\int_{ \tau( \bar{\lambda}_i )  }^{\tau_{\scaleto{\emi_i \mathstrut}{5.5pt}}  } \!\!  d\tau \,  \Eu_i (\tau)$ to \eqref{def:zint1difagain} start to be substantial? 
We can analyse this by noting that the second order approximation of $\delta z_{i}$ in $\delta \tau_{i}$ is 
\bea
\label{def:zint1difagain_2nd}
\hspace{-0.5cm} \frac{ \delta z_{i} }{ \delta \tau_{i}  }&&  \approx   
-  (\bar{z} +1 )   \braket{\Eu_i} \left(1 + \frac{1}{2} \braket{\Eu_i} \delta \tau_{i}  \right)    \, , 
\eea  
which may be used to obtain the second-order correction to \eqref{def:chidiftheory_inv}, yielding 
\bea
\label{def:chidiftheory_inv_2nd}
 \frac{1}{2}  \frac{  \sum^N_{i=1}  \sum^N_{j=1} (\text{COV}^{-1})_{ij}     \braket{\Eu_i}^2   ( \delta \tau_i )^2 \delta z_j     }{   \sum^N_{i=1}  \sum^N_{j=1} (\text{COV}^{-1})_{ij}     \delta \tau_i \delta z_j    } \, , 
\eea 
to be added to the right-hand-side of \eqref{def:chidiftheory_inv}. Since the sum in the numerator is over $( \delta \tau_i )^2 \delta z_j$ which is an odd power of the distance separation of the cosmic chronometer, we expect the sum to partly cancel. 
The level of cancelation depends on the selection function of the survey, but for catalogues with approximate distributional symmetry around $\delta \tau_i = 0$, we expect a high degree of cancelation of \eqref{def:chidiftheory_inv_2nd}, such that the leading order expression in \eqref{def:chidiftheory_inv} may for instance be expected to be reliable at the percent level, even when $\braket{\Eu_i} \delta \tau_{i}$ is several percent of magnitude.

\section{Discussion} 
We have provided results for the differential age signal for the very broad class of cosmological space-times described in section~\ref{sec:setup}.  
The main result is given in \eqref{def:chidiftheory_inv}, which provides a kinematic interpretation of a realistic estimator of the differential age signal. 
Specifically, the differential age signal probes (weighted averages of) the line-of-sight average $\braket{\Eu_i}$, which in turn measures the projected expansion rate along the null-line segment.  
Numerical investigations of a variety of non-trivial statistically homogeneous and isotropic space-times \cite{Lavinto:2013exa,Koksbang:2017arw,Koksbang:2020qry} show that $\Eu$ can, with a high level of accuracy, be replaced by (one third of) the average volume expansion rate in the matter frame of the emitting sources if the null-line segment considered is large compared to the largest structures in the models. 
The main result of this paper, combined with these analyses, shows that the differential age signal is indeed a robust probe of the volume average expansion rate of the Universe for cosmological space-times when the volume spanned by the survey is comparable to or larger than a suitable homogeneity scale of the space-time.   

This property of the differential age signal is not obvious once we go beyond the FLRW metrics. This was pointed out in \cite{Koksbang:2021qqc,Koksbang:2022ari}, where it was shown in a concrete toy model that neither the redshift drift signal nor the dipole in the distance--redshift relation are robust probes of the volume-average expansion rate once one considers non-trivial space-time scenarios with spatial curvature or backreaction, while it was conjectured that cosmic chronometers would correctly trace the expansion rate. 
The derivations of this paper give theoretical support to this conjecture in a general context. 
The statistical scatter and bias of \eqref{def:chidiftheory_inv} relative to the mean expansion-rate due to structures may be analysed within a specified model of the cosmic structures and a specified survey by application of the formula \eqref{def:chidiftheory_inv} or a modification of this corresponding to the statistical estimator of interest. 

The redshift drift signal is often mentioned as a direct probe of the cosmic expansion rate, which is true within FLRW models, but does not generally hold for other space-time metrics such as, for instance, the Lema\^itre-Tolman-Bondi models \cite{Koksbang:2022upf}. The failure of redshift drift to probe the expansion rate directly comes from a correction term in the formula for redshift drift for a general universe-model that prevents the naive generalisation of the FLRW redshift drift signal, $\dot{z} = (1+z) H_\obs - H_\emi$ with $H \rightarrow \Eu$, to be valid; cf. eq. (1) of \cite{Heinesen:2021nrc}. 
This paper shows that the differential age signal \emph{can} on the contrary be generalied in a relatively straight forward way from the FLRW expression $\delta z / \delta t = - (1+z) H$, with $H  \rightarrow \braket{\Eu}$ where $\braket{\Eu}$, cf. \eqref{def:zint1dif}, is the line-of-sight average of the expansion variable $\Eu$ along the null-line segment corresponding to the age increment $\delta t$. 
While distance--redshift cosmography may be used to extract the regional expansion rate model-independently \cite{Heinesen:2020bej,Maartens:2023tib} (some care must however be taken in interpreting the signal for an anisotropic coverage of the sky of the survey \cite{Macpherson:2024zwu}), it is not ideal for model-independently determining the expansion rate at higher redshifts, where a dynamical model is needed for interpreting the distance--redshift graph. 
In light of the Hubble tension it is particularly valuable to have cosmic chronometers as independent direct probes of the expansion rate \cite{Busti:2014dua,Gomez-Valent:2018hwc}. 
We argue, based on this work, that cosmic chronometer data is going to be a really valuable model-independent probe of the expansion rate of the Universe. 

The biggest challenge in exploring the theoretical advantages of the differential age signal are the large systematic errors associated with calibration of the ages of the chronometers. 
However, it may be possible to bring the total error budget of the differential age signal down to below 1 percent with future measurements by improving the stellar population synthesis models that are used in the calibration \cite{Moresco:2020fbm}.

\section{Conclusion} 
\label{sec:conclusion} 
We have formulated the differential age signal in cosmological space-times, using only a very limited set of assumptions listed in section~\ref{sec:setup}.  
When the space-time admits a congruence of cosmic chronometers with ages, $\tau$, and measured redshifts, $z$, we get an expression for the differential age signal $\delta z / \delta \tau$ (see \eqref{def:zint1dif}) that generalises the FLRW result for a single pair of chromometers.  
Using this result, we have derived the expression for an estimate of the differential age signal obtained from multiple cosmic chronometers (see \eqref{def:chidiftheory_inv}). 
Our analysis shows that large catalogues of cosmic chronometers are expected to probe the average expansion rate of the Universe accurately, even in space-times where this is not expected by other probes (such as redshift drift) of the cosmic expansion rate. 
Our formulae provide a framework for quantifying estimates of the differential age signal made in specified models for cosmic structure formation.

\vspace{6pt} 
%%%%%%%%%%%%%%%%%%%%%%%%%%%%%%%%%%%%%%%%%%%%%%%%%%%%%%%%%%%%%%%%%%%
\begin{acknowledgments}
%%%%%%%%%%%%%%%%%%%%%%%%%%%%%%%%%%%%%%%%%%%%%%%%%%%%%%%%%%%%%%%%%%%
I thank Sofie Koksbang for useful discussions. AH is funded by the Carlsberg foundation. 
\end{acknowledgments}

\bibliographystyle{mnras_unsrt}
\bibliography{refs}

\end{document}